# Features of Propagation of Light in the Linear Array of Dielectric Spheres

**Michael Gozman**[+*], **Ilya Polishchuk**[*$+], **Alexander Burin**[+]
[+]*Department of Chemistry, Tulane University, New Orleans*
[*]*Max Planck-Institute für Physik Komplexer Systeme, D-01187 Dresden, Germany*
[$]*RRC Kurchatov Institute, Kurchatov Sq 1, 123182 Moscow Russia*
*Tel: (+49 351) 8711126, Fax: (+49 351) 871111, e-mail: iyp@mpipks-dresden.mpg.de*

**ABSTRACT**
A finite length linear chain of dielectric loss-less identical spheres is considered. A propagation of dipole radiation in the chain of particles induced by the point dipole source placed near one end of the chain is investigated. It is found that at sufficiently large refractive index there exist frequency pass bands around every low frequency Mie resonance. In particular, if the dipole oscillates across the chain axis, one can reveal a longitudinal mode frequency pass band if refractive index $n_r$ of the spheres exceeds 1.9. Then, if the dipole oscillates transversely to the chain axis, the transverse frequencies pass bands show up depending on the chain length. In this case, the pass band is formed if the length chain is large enough. Three dielectric materials with $n_r = 1.9, n_r = 2.7, n_r = 3.5$ are considered what corresponds to ZnO, $TiO_2$, GaAs. It is found that the top of the frequency pass band corresponds to the top of the Brillouin band edge in the quasi-momentum space. On the order hand, the bottom of the frequency pass band corresponds to the guiding wave criterion [2, 3]. This explains the remarkable feature of the band picture established for infinite chain in Ref. [1]: the band structure breaks down as the wavevector becomes small enough. The multisphere Mie scattering formalism is used to calculate how the amplitude of the radiation changes along the chain.

## 1. INTRODUCTION

One-dimensional periodic chains of dielectric spheres are considered for potential applications in optical waveguides. Along with a high quality factor, such arrays can be used to control the velocity of optical waves [3]. Investigation of a travelling optical mode in an infinite linear periodic array of dielectric spheres is equivalent to the solution to the Maxwell equations with permittivity and permeability changing periodically, see Fig. (1). In this case, one should expect a strong similarity between the Maxwell equations describing the chain under consideration and the one-dimensional Schrödinger equation in periodic potential. Any solution to the Schrödinger equation is characterized by the quasiwave vector $|k| \le \pi / a$, where $\boldsymbol{a}$ is the space period.

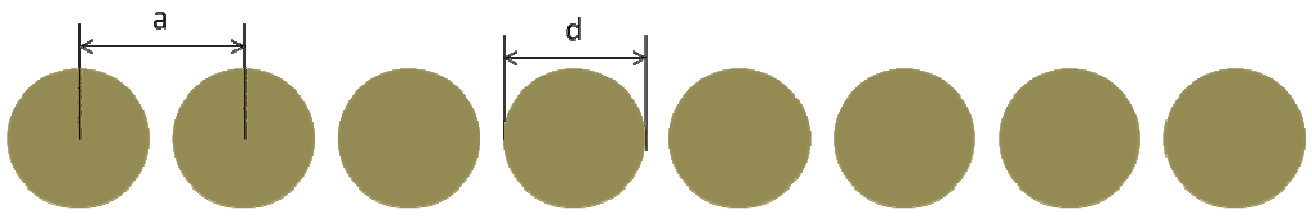

*Figure 1. Linear chain of scattering particles.*

The energy spectrum is a sequence of interlaced allowed and forbidden bands. Within each allowed band every wavevectors $\boldsymbol{k}$ from the interval $(-\pi / a, \pi / a)$ gives rise to a certain solution. One can naively suppose that the same picture should take place if one considers the Maxwell equations for the linear array mentioned above. Instead, it occurs that the photonic band structure reveals the unusual peculiarity.

Let the travelling wave in the chain, see Fig. 1, is characterized by the spatial-time dependence $e^{i(\omega t - kz)}$ with a real frequency and a real quasi-wavevector with a certain dispersion law $\omega(k)$; here and in what follows we put the vacuum speed of light $c = 1$. An explicit numerical analysis reveals that the dispersion low can be found only for the wavevectors k obeying the criterion

$$\omega(k) < |k| < \pi / a \qquad (1)$$

This result can be interpreted as follows. Consider an excitation (a polariton) with frequency $\omega$ travelling along the infinite chain. This excitation is a superposition of plane waves with momentums $k_n = k + 2\pi n / a$, $n = 0,1,2....$ The excitation can escape from the chain, being transformed into a free photon. Because of the energy conservation law, the frequency of the photon is $\omega$. The momentum of this photon $p = \omega$. This photon originates from the superposition of the excitations with the momentums $k_n$. Because of the momentum conservation, the projection of the photon momentum $p$ on the chain axis is $p_{\|} > k$. Therefore, $k < p_{\|} < p = \omega$. If this is not the case, the radiation of a photon is impossible. Thus, the propagation of excitation without loss on the photon excitation takes place if the guiding criterion (1) is fulfilled.

In paper [1] R. Shore and A. Yaghjian investigated low energy polariton modes in infinite chains of particles having a high refractive index within the dipole approximation. They discovered the existence of the energy bands free of radiative decay. Our goal is to extend the consideration of Shore and Yaghjian to the more realistic case of finite length chains with the polaritons excited by an external source. We demonstrate that the energy bands discovered in Ref. [1] correspond to the pass bands of finite chains where the radiation propagates practically without losses.

Note that the chains under consideration also were investigated in Refs. [4, 5]. In paper [4] the FDTD approach has been applied to investigation of the system containing a finite number of spheres. In paper [5] the multipole scattering effects up to orbital momentum $l = 80$ were considered.

In our previous publications [2, 3] we investigated the behavior of the quality factor in linear and circular arrays of dielectric lossless spheres as function of the frequency, number of particle in the array and refractive index of the spheres. The multi-sphere Mie scattering formalism (MSMS) [6, 7] was used. It was established in Refs. [2, 3] that each Mie resonance inherent in a sphere gives rise to the modes having extremely high quality factor. In the present paper we consider a finite chain of dielectric spheres and investigate how the electromagnetic wave propagates along the chain. For this purpose, a point dipole source (either electric or magnetic one) is placed onto the axis z near one of the end of the finite chain (see Fig. (1)). The main goal is to investigate how this radiation propagates along the chain as the dipole frequency changes. Then, we compare how the results obtained are related to those of Refs. [2, 3]. We expect that the low frequency high-quality modes found in Refs. [2, 3] correspond to the pass bands found in these paper.

## 2. The Multisphere Mie Scattering Formalism.

The multisphere Mie scattering formalism [6, 7] uses the vector spherical wave function as a basis for expansion of solutions to the Maxwell's equations in the frequency domain. In the dipole approximation the scattering of the excitation by the linear array of spheres is described by the system of equation

$$\frac{a_{m1}^{\ l}}{\bar{a}_1} + \sum_{j\neq l}^{(1,N)} (A^{jl}{}_{m1m1} a_{m1}{}^{l} + B^{jl}{}_{m1m1} b_{m1}{}^{l}) = p^{l}{}_{m1}, \qquad (2)$$

$$\frac{b_{m1}^{\ l}}{\bar{b}_1} + \sum_{j\neq l}^{(1,N)} (B^{jl}{}_{m1m1} a_{m1}{}^{l} + A^{jl}{}_{m1m1} b_{m1}{}^{l}) = q^{l}{}_{m1}. \qquad (3)$$

Here $\bar{a}_1, \bar{b}_1$ are the dipole Mie scattering coefficients of the spheres, $A^{jl}{}_{m1m1}$ and $B^{jl}{}_{m1m1}$ are vector translation coefficients describing dipole-dipole interaction between the spheres, $p^{l}{}_{m1}$ and $q^{l}{}_{m1}$ are the expansion coefficient for the incident dipole radiation, $a^{l}{}_{m1}$ and $b^{l}{}_{m1}$ are the expansion coefficients for the scattered radiation; $m = 0, \pm 1$ is the projection of the photon orbital momentum, $l$ and $j$ stand for the number of the sphere. If $m = 0$, the vector translation coefficients $B^{jl}{}_{m1m1}$ vanish, and equations (2) and (3) become uncoupled. In particular, in the case of the source is a magnetic vibrating dipole, $p^{l}{}_{m1} = 0$, while $q^{l}{}_{m1} = A^{0l}{}_{0101}$. In this case all $a_{01}{}^{l} = 0$ while $b_{01}{}^{l}$ obey the equation

$$\frac{b_{01}^{\ l}}{\bar{b}_1} + \sum_{j\neq l}^{(1,N)} A^{jl}{}_{0101} b_{01}{}^{l} = A^{0l}{}_{0101}\,. \qquad (4)$$

## 3. Propagation of the dipole radiation on the linear chain

We analyze the guiding properties of the chain of particle with refractive indexes $n_r = 3.5, 2.7, 1.9$ corresponding to $\mathrm{GaAs}, \mathrm{TiO_2}, \mathrm{ZnO}$. The chain is supposed to consist of $N = 200$. In addition, we consider the case of touching spheres $a = 2d$. In this case guiding criterion (1) is satisfied the most easily [2, 3]. To analyze the propagation of the dipole excitation, one should solve system of equations (2)-(3). If the dipole vibrates transversely to the $z$-direction, one should put $m = 1$ in these equations. The case $m = -1$ is the same. If the dipole vibrates across the $z$- direction, one should put $m = 0$. In the last case if, in addition, the dipole is a magnetic one, we should solve equation (4).

These equations have been solved numerically using the MatLab software package. The results of calculations are presented in Fig. 2 – Fig. 7. The curves in these figures illustrate the behaviour of the field amplitude at the sphere most remote from the dipole emitter. Within certain frequency intervals, this amplitude is almost independent on the frequency as well as practically independent on the total number of spheres (we have verified this for $N = 300, 400$ in GaAs). These frequency intervals should be associated with frequency pass bands. One can see that for longitudinal modes the frequency pass bands exist for GaAs and $TiO_2$ both for electric and magnetic dipole radiation. For ZnO, there are no longitudinal travelling modes at all. These conclusions correspond to the results obtained in Refs [2, 3]: the high quality factor longitudinal mode exist if $n_r > 2$. In particular, the magnetic longitudinal travelling modes exist in GaAs within the frequency interval $0.86 < \omega < 0.94$, while the electric longitudinal travelling modes exist in GaAs within the frequency interval $1.18 < \omega < 1.2$. These values of edges of the frequency pass bands obtained allow explicit physical interpretation.

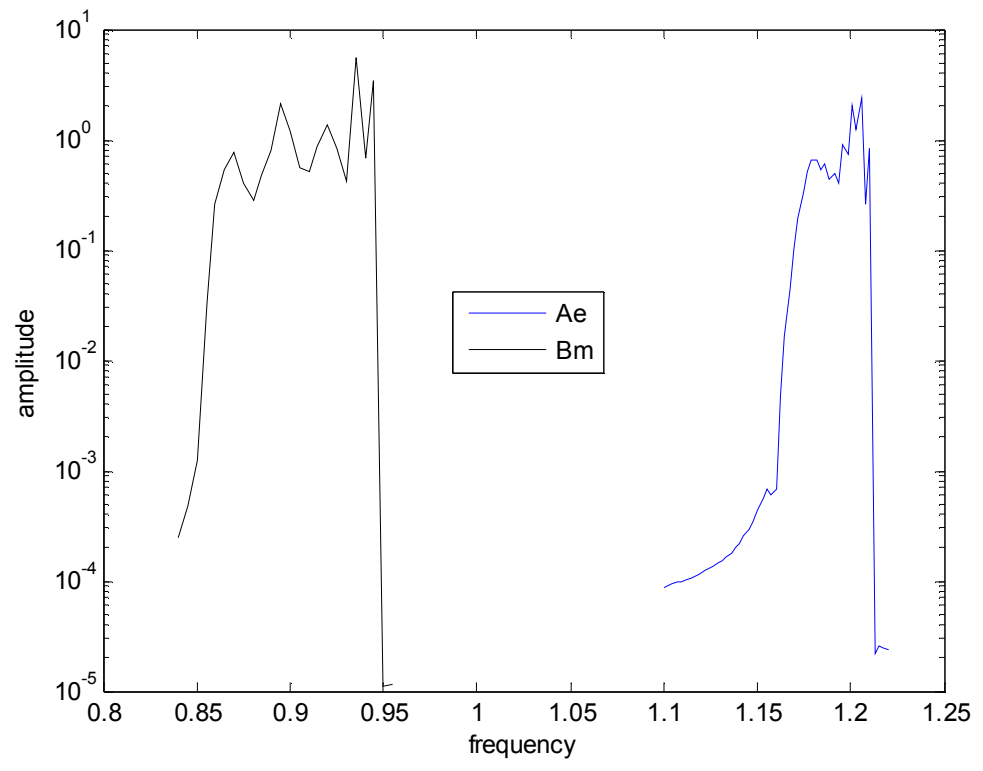

*Figure 2. The partial amplitude of the longitudinal modes on the 200$^{th}$ sphere of GaAs.*

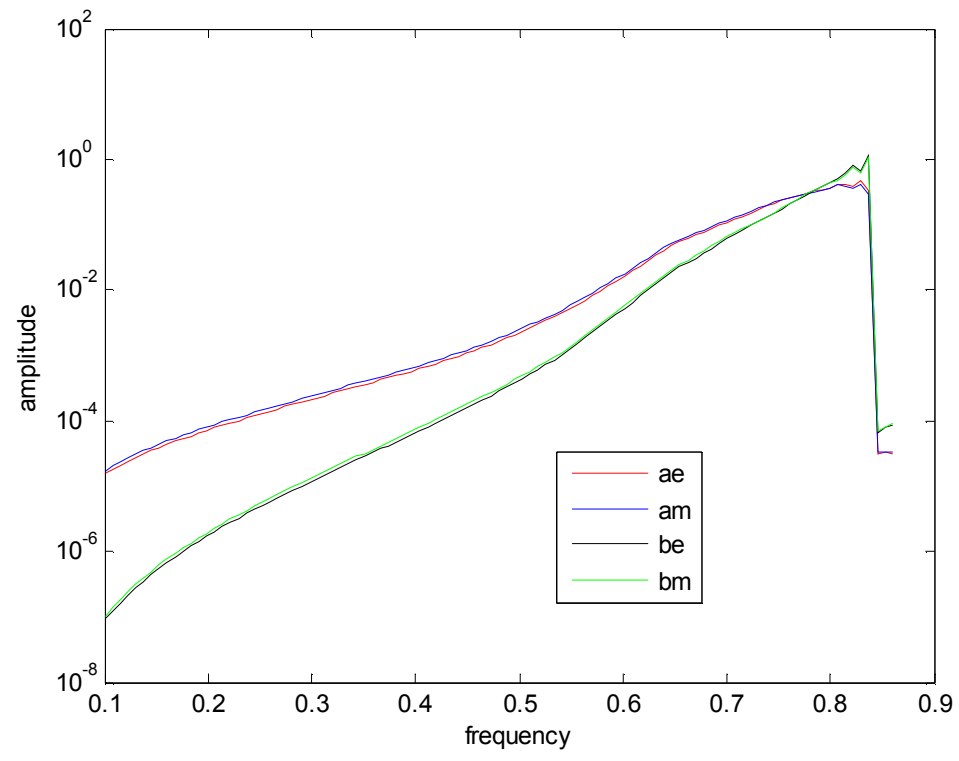

*Figure 3. The partial amplitude of the transverse modes on the 200$^{th}$ sphere of GaAs.*

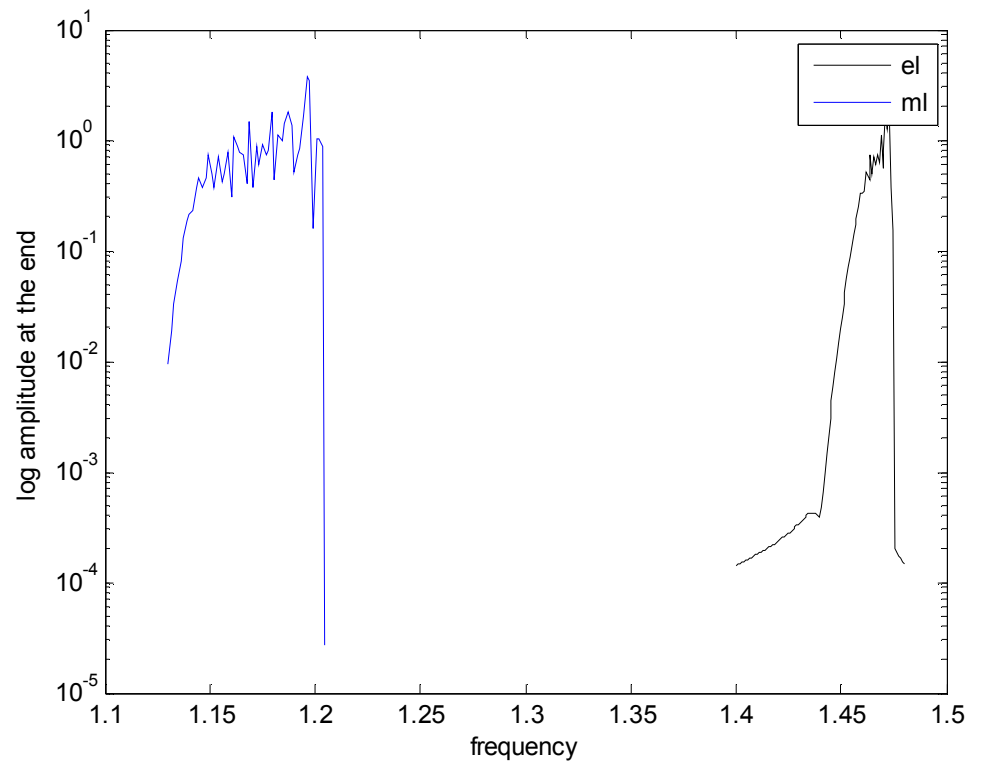

*Figure 4. The partial amplitude of the longitudinal modes on the 200$^{th}$ sphere of $TiO_2$.*

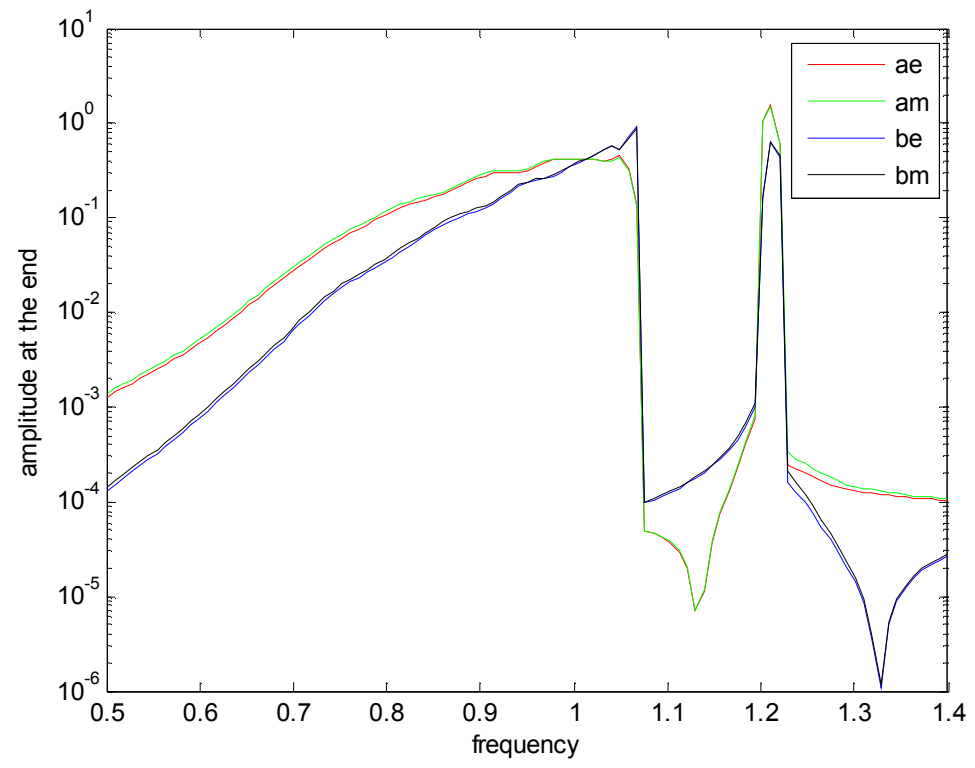

*Figure 5. The partial amplitude of the transverse modes on the 200$^{th}$ sphere of $TiO_2$*

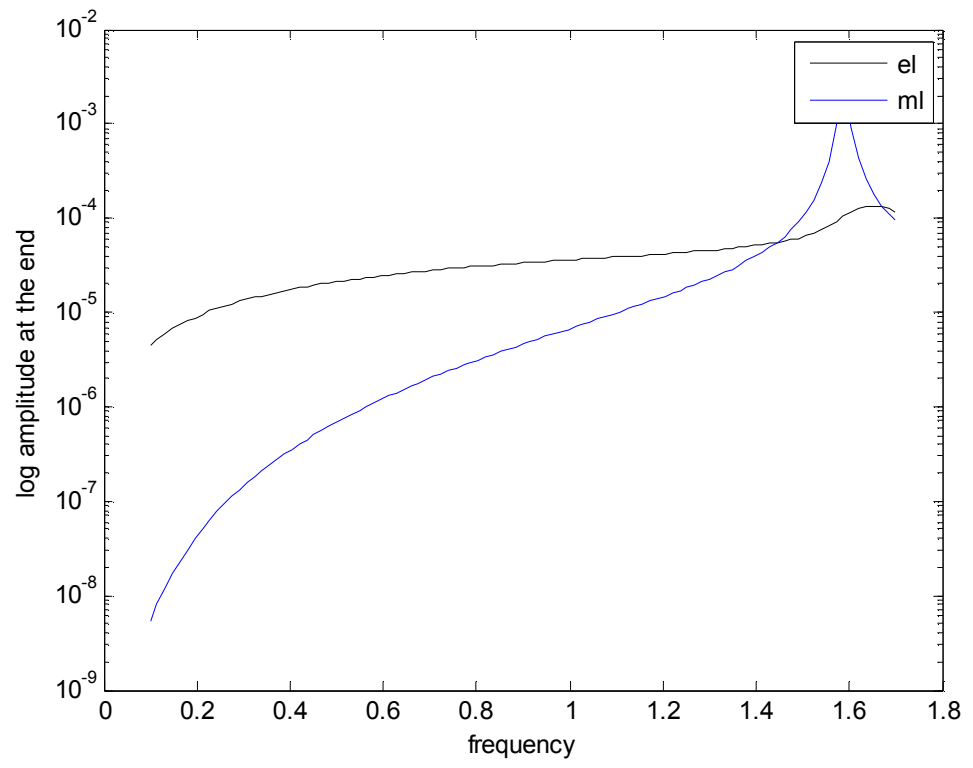

*Figure 6. The partial amplitude of the longitudinal modes on the 200$^{th}$ sphere of ZnO.*

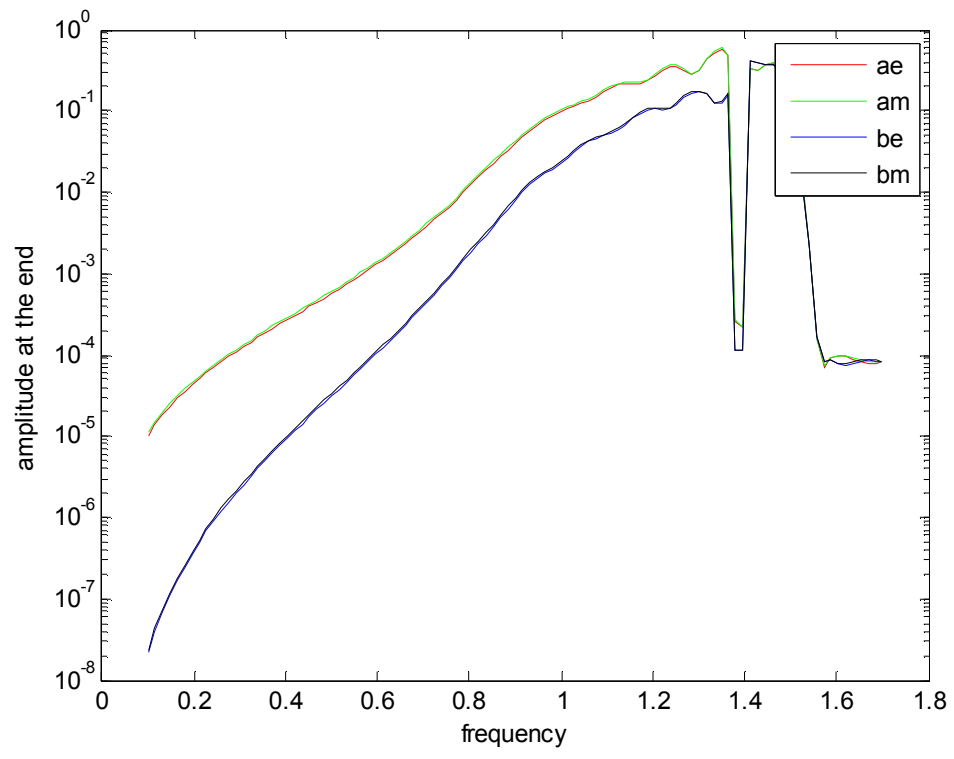

*Figure 7. The partial amplitude of the transverse modes on the 200$^{th}$ sphere of $TiO_2$*

For the certainty, consider the magnetic longitudinal dipole mode in GaAs in the infinite periodical chain. In this case, each propagating mode is characterized by the quasiwave vector $|k| < \pi / a$. Using approach developed in Ref. [2, 3] for solving Equation (4), one can show that for $k = \pi / a$ the corresponding frequency equals to 0.94. This value coincides with the left edges of the black curve in Fig. 2. Let then one gradually decreases the quasiwave vector. It occurs, that for wavevectors $|k| < 0.86$ there is now real frequency $\omega$. Thus, in the infinite chain the spectrum brakes inside the Brillouin zone. Let us note that this value coincides with the low edge of the frequency pass band in Fig. 2. It is important to pay attention that this coincidence corresponds to guiding criterion (1). This feature is inherent in any longitudinal mode if $n_r > 2$. In particular, this takes place also in $TiO_2$ and does not take place in ZnO.

On the other hand, as it is illustrated by Figs. 3, 5, 7 the transverse travelling modes show up for the long enough chain. This completely corresponds to the results of Refs. [2, 3] where the physical explanation of this feature has been presented.

## 4. CONCLUSIONS

In this paper we studied the propagation of the dipole radiation along the finite linear chain of lossless dielectric spheres with different refractive index. It is found that guiding transverse mode show up if the chain is long enough, while the longitudinal guiding modes exist only if refractive index $n_r > 2$. These features also were marked to in Refs. [2, 3]. Each frequency pass band are created by one of the Mie resonance. The upper border of each pass band corresponds to the top of the Brillouin band of the infinite chain, while the lower border of the pass band is connected with guiding criterion (1). This means that at small enough wavevectors the frequency spectrum of eigen modes in the chain becomes unstable with respect to emission of a free photon. There is no difference between the features of propagation both transverse magnetic and electric dipole radiation. The simulation of our paper and the physical interpretation, in particular, are supported by the numerical results obtained in Ref. [1].

## ACKNOWLEDGEMENTS

This work is supported by the U.S. Air Force Office of Scientific Research (Grant No. FA9550-06-1-0110 ) and by Russian Fund for Basic Research (Grant No 07-02-00309).